\documentclass[paper,11pt]{article}
\pdfoutput=1

\usepackage[pdftex,bookmarks,colorlinks]{hyperref}
\usepackage[pdftex]{graphicx}
\usepackage[amssymb]{SIunits}
\usepackage{mathrsfs,mathcomp}
\usepackage{fancyhdr}
\fancypagestyle{plain}{
\fancyhead[CO]{\footnotesize{Fluid dynamics video for the 30th Annual Gallery of Fluid Motion\\ 65th Annual Meeting of the American Physical Society, Division of Fluid Dynamics \\San Diego, CA, Nov 2012. Entry \#84160}}
}

\begin{document}

\title{Diving with microparticles in acoustic fields }

\author{\'Alvaro Mar\'in$^{1}$\footnote{alvarogum@gmail.com}, Massimiliano Rossi$^{1}$, Rune Barnkob$^{2}$, Per Augustsson$^{3}$, \\
Peter Muller$^{2}$, Henrik Bruus$^{2}$,Thomas Laurell$^{3}$, Christian K\"ahler$^{1}$\\
\emph{$^1$Universit\"{a}t der Bundeswehr M\"{u}nchen, Germany}\\
\emph{$^2$Technical University of Denmark, DTU Physics, Denmark}\\
\emph{$^3$Lund University, Sweden}\\}

\maketitle

\begin{abstract}

Sound can move particles. A good example of this phenomenon known as \emph{acoustophoresis} is the Chladni plate, in which an acoustic wave is induced in a metallic plate and particles migrate to the nodes of the acoustic wave. For several years, acoustophoresis has been used to manipulate microparticles in microscopic scales\cite{Bruus2011,Laurell2007}. In this fluid dynamics video, submitted to the 30th Annual Gallery of Fluid Motion, we show the basic mechanism of the technique and a simple way of visualize it. Since acoustophoretic phenomena is essentially a three-dimensional effect, we employ a simple technique to visualize the particles in 3D. The technique is called Astigmatism Particle Tracking Velocimetry \cite{APTV} and it consists in the use of cylindrical lenses to induce a deformation in the particle shape, which will be then correlated with its distance from the observer. With this method we are able to \emph{dive} with the particles and observe in detail particle motion that would otherwise be missed. The technique not only permits visualization but also precise quantitative measurements that can be compared with theory and simulations\cite{runeAPS, massiAPS}. 
\end{abstract}

\section{Experimental Details}

The Chladni plate experiment was build and filmed in the Technical University of Denmark, DTU Physics by using a metallic plate, a wave generator and sugar. 

The microfluidic channel shown (from Thomas Laurell group in Lund University) and used for the acoustophoretic experiments corresponds to a rectangular cross section channel ($L$ = 35 mm, $w$ = 377 $\upmu$m, and $h$ = 157 $\upmu$m) was etched in silicon. A Pyrex lid was anodically bonded to seal the channel. The outer dimensions of the chip are $L$ = 35 mm, $W$ = 2.52 mm, and H = 1.48 mm. From top and down, glued together the chip was placed on top of a piezoceramic transducer (piezo), an aluminum slab to distribute heat evenly along the piezo, a Peltier element to enable temperature regulation based on readings from a temperature sensor placed near the chip on the transducer, and an aluminum heat sink. Ultrasound vibrations were generated in the piezo by applying an amplified sinusoidal voltage from a function generator, and the resulting piezo voltage was monitored using an oscilloscope. More details as well as sketches and figures of the device can be found in Ref.~\cite{Augustsson2011}.

Spherical polystyrene particles with diameters of 5.33~$\upmu$m (SD 0.09) and 0.537~$\upmu$m (PDI 0.005) fabricated by Microparticles GmbH were used for the experiments. All particles were labeled with a proprietary fluorescent dye to be visualized using an epifluorescent microscopic system. The images were taken using a 12-bit, 1376$\times$1040 pixels, interline transfer CCD camera (Sensicam QE, PCO GmbH) connected to an epifluorescent microscope (DM2500 M, Leica Microsystems CMS GmbH, Wetzlar, Germany). A principal objective lens with 20$\times$ magnification and 0.4 numerical aperture was used. An additional cylindrical lens with focal length $f_\mathrm{cyl} = 150$~mm was placed in front of the camera CCD to obtain the astigmatic particle images as described in~\cite{APTV}. With this configuration a measurement volume of 900$\times$600$\times$120~$\upmu$m$^3$ was obtained so that a two-position scan along the $z$-direction was necessary to cover the whole cross-sectional area of the channel with the desired spatial resolution and measurement precision. The illumination was provided by a continuous diode-pumped laser with 2~W at 532~nm wavelength. 

Both 3D trajectories of the 5-$\upmu$m and the 0.5-$\upmu$m particles were recorded with the piezo operated at 1.94~MHz, which generates a half wave with a node at the center of the channel.

Finally, please note that all 3D animations in the video correspond to real experimental of microparticles measurements. 


\bibliographystyle{plainnat}

\end{document}